\documentclass[aps,prl,twocolumn,nofootinbib,a4paper,amsfonts,amssymb,amsmath,floatfix,superscriptaddress]{revtex4-2}
\usepackage{graphicx}
\usepackage{xcolor}
\usepackage{dsfont}
\usepackage{physics}

\usepackage[T1]{fontenc}
\usepackage[utf8]{inputenc}

\newcommand{\beq}{\begin{equation}}
\newcommand{\eeq}{\end{equation}}
\newcommand{\beqa}{\begin{eqnarray}}
\newcommand{\eeqa}{\end{eqnarray}}

\usepackage{hyperref}
\usepackage{amsmath,amsfonts}

\def\opone{\leavevmode\hbox{\small1\normalsize\kern-.33em1}}
\def\C{{\cal C}}
\def\cb{\tiny{\ket{0}}}
\def\sb{\tiny{\ket{\varphi}}}

\begin{document}

\title{Iso-entangled bases and joint measurements}
\author{Flavio Del Santo}
\affiliation{Group of Applied Physics, University of Geneva, 1211 Geneva 4, Switzerland}
\affiliation{Constructor University, Geneva, Switzerland}

\author{Jakub Czartowski}
\affiliation{Doctoral School of Exact and Natural Sciences, Jagiellonian University, ul. Łojasiewicza 11, 30-348 Kraków, Poland}
\affiliation{Faculty of Physics, Astronomy and Applied Computer Science, Jagiellonian University, ul. Łojasiewicza 11,
30-348 Kraków, Poland}

\author{Karol \.Zyczkowski}
\affiliation{Faculty of Physics, Astronomy and Applied Computer Science, Jagiellonian University, ul. Łojasiewicza 11,
30-348 Kraków, Poland}
\affiliation{Center of Theoretical Physics, Polish Academy of Sciences, al. Lotników 32/46 02-668 Warszawa}

\author{Nicolas Gisin}
\affiliation{Group of Applied Physics, University of Geneva, 1211 Geneva 4, Switzerland}
\affiliation{Constructor University, Geneva, Switzerland}

\begin{abstract}
\noindent
     While entanglement between distant parties has been extensively studied, entangled measurements have received relatively little attention despite their significance in understanding non-locality and their central role in quantum computation and networks. We present a systematic study of entangled measurements, providing a complete classification of all equivalence classes of iso-entangled bases for projective joint measurements on 2 qubits. The application of this classification to the triangular network reveals that the Elegant Joint Measurement, along with white noise, is the only measurement resulting in output permutation invariant probability distributions when the nodes are connected by Werner states. The paper concludes with a discussion of partial results in higher dimensions.
   \vspace{1cm}
\end{abstract}

\date{\small \today}
\maketitle

\section{Introduction}
\medskip
In 1935, Schr{\"o}dinger stated that entanglement is not one but rather \textit{the} characteristic
trait of quantum mechanics. Indeed, today it is well known that entanglement is not only necessary for the violation of celebrated Bell inequalities---disproving local hidden variables---but for  most of  the applications in quantum information science such as security proofs of quantum cryptography or quantum
teleportation, to name but a few examples.
\medskip

Entanglement is sometimes called the “quantum teleportation channel”. However, this overlooks
the fact that entanglement plays a dual role in this fascinating process: first as the channel
connecting the distant parties, indeed, but also in the joint measurement that triggers the
teleportation process~\cite{bennett1993teleporting}. Similarly, these joint measurements are at the heart of entanglement swapping~\cite{zukowski1993event} and dense coding~\cite{bennett1992communication}. Formally, they are represented in quantum theory by self-adjoint operators which in turn are characterized by their eigenvectors. When these eigenvectors are entangled, one says that the measurement is
entangled. For example, in the best known joint measurement, the eigenvectors are the Bell states
which are all maximally entangled.
\medskip

Entanglement between distant parties, traditionally named Alice and Bob, is by now well-studied and understood. However, entangled measurements received so far relatively little attention~\cite{jozsa2003entanglement, chitambar2014everything}
 and have never  been studied in a systematic manner. This  is somewhat surprising and disappointing, given their central role in quantum computation~\cite{nielsen2002quantum}  and quantum networks~\cite{tavakoli2022bell}. In fact, it has been recently pointed out that understanding  entangled measurements is one of the most interesting future directions in the foundations of quantum physics~\cite{cavalcanti2023fresh}. 

Studying entanglement beyond maximal value can lead to novel understanding and applications. Indeed, it is known by now that maximally entangled states are not always the best resource for quantum information tasks: non-maximally entangled quantum states in general outperform maximally entangled ones in most measures of non-locality, such as Bell inequalities, entanglement simulation with communication, the detection loophole and quantum cryptography~\cite{methot2006anomaly, brunner2005entanglement}. While this has not been investigated nearly as thoroughly for joint measurements, it has been shown that non-maximally entangled measurements represent stronger resources for certain tasks, such as the violation of bilocality ~\cite{tavakoli2021bilocal}.

In this paper, we  provide the first systematic study of entangled measurements for the simplest case. The problem is known to be difficult in full generality, hence we assume that all the eigenvectors have the same degree of entanglement, i.e., they form an iso-entangled basis (previous works on non-maximally entangled joint measurements and iso-entangled bases are~\cite{koniorczyk2005nonmaximally, karimipour2006equientangled, rajchel2018robust, tavakoli2021bilocal, Czartowski2021bipartitequantum, huang2022entanglement, pimpel2023entangled}). Moreover, we mostly limit our analysis to projective joint measurements on 2~qubits.



Here, we give a complete classification of all iso-entangled bases of 2 qubits, up to the natural equivalence relation of local unitary rotations and swapping of the qubits. Next, we apply our
parametrization to the triangular network and prove that the Elegant Joint Measurement (and white
noise) is the only measurement that leads to output permutation invariant probability distributions when the nodes are connected by identical Werner states. Finally, we discuss partial results in higher dimensions.


\section{Complete classification of all equivalence classes of iso-entangled bases of 2 qubits}
Consider measurements on two qubits, i.e., the partition of the Hilbert space  $\C^4=\C^2\otimes\C^2$ is fixed. An iso-entangled basis is an orthonormal basis s.t. all 4 vectors $\ket{\psi_j}$, $j=1,\dots4$, have the same degree of entanglement.

There are many measures of entanglement, but for pure bipartite states $\rho_{AB}=\ket{\psi}\bra{\psi}_{AB}$ they are all equivalent~\cite{vidal2000entanglement}. We quantify the degree of entanglement by its {\sl tangle}, equal to squared concurrence~\cite{wootters1998entanglement, rungta2001universal, peters2004mixed}
\beq\label{conc}
\xi=2\left(1-\textrm{Tr}(\rho_A^2)\right)\in[0,1],
\eeq
where $\rho_A=\textrm{Tr}_B(\rho_{AB})$ is the reduced density matrix; this monotonically quantifies entanglement from 0 (separable states) to 1 (maximally entangled states).



\vspace{0.2cm}
\textbf{Defintion 1} (Local equivalence of bases) Let us define the equivalence relation $\sim$ : two bases $B_1$ and $B_2$ are equivalent iff they are identical under local unitaries, $U_i$ (equivalently, local changes of basis), or identical under swap $S_{A\leftrightarrow B}$ and local unitaries, i.e.
\begin{equation} \label{eq:equivalence}
	B_1 \sim B_2\Leftrightarrow B_2 = (U_A\otimes U_B)(\cdot) B_1 P,
\end{equation}
 with $(\cdot) \in \{\mathds{1},\textrm{S}_{A\leftrightarrow B}\}$ and $P$ an arbitrary permutation.
\vspace{0.2cm}


Our goal is to find a parametrization of each family of equivalence classes. Starting with 12 real parameters for an arbitrary dephased orthonormal basis of $\C^4$, we subtract $3+3$ parameters for local changes of bases, and the 3 constraints that all 4 vectors have the same degree of entanglement. We thus expect an iso-entangled basis of two-qubits to depend in general on 3 parameters.

Our main result consists in the following proposition:

\vspace{0.2cm}
\textbf{Proposition 1} (Complete classification of iso-entangled bases of 2 qubits) 
All equivalence classes of iso-entangled bases on the space $C$ with respect to the relation~\eqref{eq:equivalence}  constitute a three-dimensional manifold composed of two families, together with the closure of discontinuous submanifolds, given by three additional families of equivalence classes of smaller dimension. The specific functional form of the families is provided in Eq.~\eqref{monster} for the general family,~\eqref{bellfamily} for the Bell family, and in Eqs.~\eqref{separablefamily},~\eqref{elegantfamily} for the families of smaller dimensions. 
\vspace{0.5cm}



\subsection{Constructive proof}
Let $B$ be a matrix of order 4 whose columns are 4 basis vectors $\{\ket{\psi_1},\ket{\psi_2},\ket{\psi_3},\ket{\psi_4}\}$ in $\C^4=\C^2\otimes\C^2$. Let us write $B$ in the following skewed basis (by applying local change of basis only), consisting only of product states, but in general different from the computational basis: 
\beq\label{skewedB}
\ket{0,0},\ket{0,1},\ket{1,\varphi},\ket{1,\varphi^\perp},
\eeq
where $\ket{\varphi}=\cos(\tau)\ket{0}+\sin(\tau)\ket{1}$ and $\ket{\varphi^\perp}=\cos(\tau)\ket{1}-\sin(\tau)\ket{0}$. To simplify the derivation, we use the fact that any 2-dimensional subspace of $\C^4$ contains at least one product state~\cite{sanpera1998local}.
Imposing the orthonormality leads to the following parametrization of an arbitrary equivalence class of 2-qubit orthonormal bases (for derivation see SM, section A):
\beqa
\label{onbasis}
B_{\ket\varphi}=
\left(
\begin{array}{cccc}
	0& 0& -c\alpha\cdot e^{i\gamma}& s\alpha\cdot e^{i\gamma}\\
	s\delta\cdot c\theta&c\delta\cdot c\theta&-s\alpha\cdot s\theta&-c\alpha\cdot s\theta\\
	s\delta\cdot s\theta&c\delta\cdot s\theta&s\alpha\cdot c\theta&c\alpha\cdot c\theta\\
	-c\delta\cdot e^{i\beta}&s\delta\cdot e^{i\beta}&0&0
\end{array}
\right),
\eeqa 
where we have introduced the compact notation \mbox{$c\delta=\cos{\delta}$}, \mbox{$s\delta=\sin\delta$} and similarly for $c\alpha$ and $s\alpha$, and $c\theta$ and $s\theta$. The subscript $\ket{\varphi}$ indicates that the coefficients are expressed in the basis provided in Eq.~\eqref{skewedB}. As expected, this parametrization has 6 parameters: $\alpha, \delta, \theta, \gamma, \beta$ and $\tau$ (with $\tau$ included implicitly in skewed basis~\eqref{skewedB}). 

By computing the tangle  $\xi_j$ (Eq.~\eqref{conc}) for each state $\ket{\psi_j}$ of $B$, we can now impose the constraints of iso-entanglement:
\beq\label{isoent}
\xi_i=\xi_j, \ \ \ \forall i,j\in \{1,2,3,4\}.
\eeq
Note that only 3 of these equations are independent, thus solving these constraints will lead to a parametrization depending on $6-3=3$ parameters. 
The previous equations yield the following complete set of solutions:
\begin{itemize}
    \item[(i)] $\cos \theta =0$, or
    \item[(ii)] $\sin 2\theta\neq 0$, and $\sin\tau =0 \implies \alpha = \frac{\pi}{4} =\pm \delta+\frac{l \pi}{2}$, or
    \item[(iii.a)] $\sin \theta=0$, and $\sin\tau \neq 0 \implies \alpha=\pm \delta+\frac{k \pi}{2}$, or
    \item[(iii.b)] $\cos\tau =0 \implies \alpha=\pm \delta+\frac{m \pi}{2}$, or
    \item[(iv)] $\cos\theta \neq 0$, and $\sin\theta \neq 0$, and $\cos \tau \neq 0$, and $\sin \tau \neq 0$, and $\sin (2\delta) \neq 0$, and $\sin (2\alpha) \neq 0  \implies \alpha=\pm \delta+\frac{n \pi}{2}$.
\end{itemize}
As we shall see, the first solution ($\cos \theta =0$) is somehow trivial, for it leads to all four basis states to be separable. All the other solutions imply that  $ \alpha=\pm \delta$ (omitting here the periodicity of $\pi/2$). This condition can thus be substituted in (three of) the equations~\eqref{isoent}, leading to the following simplified expressions for the iso-entanglement conditions: 
\scriptsize
\beqa
0=\xi_1-\xi_2&=&-8\cos^2\theta\sin\theta\cos\tau\cdot \label{xi1m2}\\
&&\cdot \bigl(\cos\tau\sin\theta\cos(2\delta)-\sin\tau\sin(2\delta)\cos\beta\bigr) \nonumber\\
0=\xi_3-\xi_4&=&-8\cos^2\theta\sin\theta\cos\tau\cdot \label{xi3m4}\\
&&\cdot\bigl(\cos\tau\sin\theta\cos(2\delta)+\sin\tau\sin(2\delta)\cos\gamma \bigr) \nonumber\\
0=\xi_1-\xi_3&=&8\cos^2\theta\sin\theta\cos\tau\cdot \label{xi1m3}\\
&&\cdot \sin(2\delta)\sin^2\delta\sin\tau\bigl(\cos\beta+\cos\gamma\bigr).  \nonumber
\eeqa
\normalsize
We are now in position to fully characterize the different classes of parametrizations of iso-entangled bases of two qubits. These correspond to the 5 different solutions (i)-(v) above of Eqs.~\eqref{xi1m2},\eqref{xi3m4},\eqref{xi1m3}. Note, however, that two of the solutions (namely, (iii.a) and (iii.b)), lead to equivalent families up to a swap (so they belong to the same equivalence class). Therefore, we arrive at 4 families of iso-entangled bases. We will denominate the different families $I^{(j)}$ with $j\in(1,\hdots,4)$, and we will express them in either the computational basis or in the skewed basis~\eqref{skewedB}; we will indicate this by a subscript $\ket{0}$ or $\ket{\varphi}$, respectively. We will see that each of them is characterized not only by a different functional form of the states (which reflects different geometrical properties thereof) but also by the amount of parameters which the  degree of entanglement $\xi^{(j)}$ depends on. 

\subsection{Four inequivalent families of iso-entangled bases}

Solutions (i-iv) lead to the following four families of isoentangled bases:
\vspace{0.2cm}

\noindent\textbf{1. \ Skewed product family}
\vspace{0.2cm}


Starting from condition (i), the parametrisation can be reduced to
\footnotesize
\beqa \label{separablefamily}
I^{(1)}_{\cb}=
\left(
\begin{array}{cccc}
	1 & 0 & 0 & 0 \\
 0 & 1 & 0 & 0 \\
 0 & 0 & \cos\tau & -\sin\tau \\
 0 & 0 & \sin\tau & \cos\tau
\end{array}
\right).
\eeqa 
\normalsize
where the other parameters have 
been absorbed into local transformations.
Note, that the degree of entanglement is  $\xi^{(1)}=0$, independently of $\tau$.
As already mentioned, this family contains only product bases, equivalent to skewed basis provided in Eq.~\eqref{skewedB}. 

\begin{figure}[h]
    \centering
    \includegraphics[width=.4\textwidth]{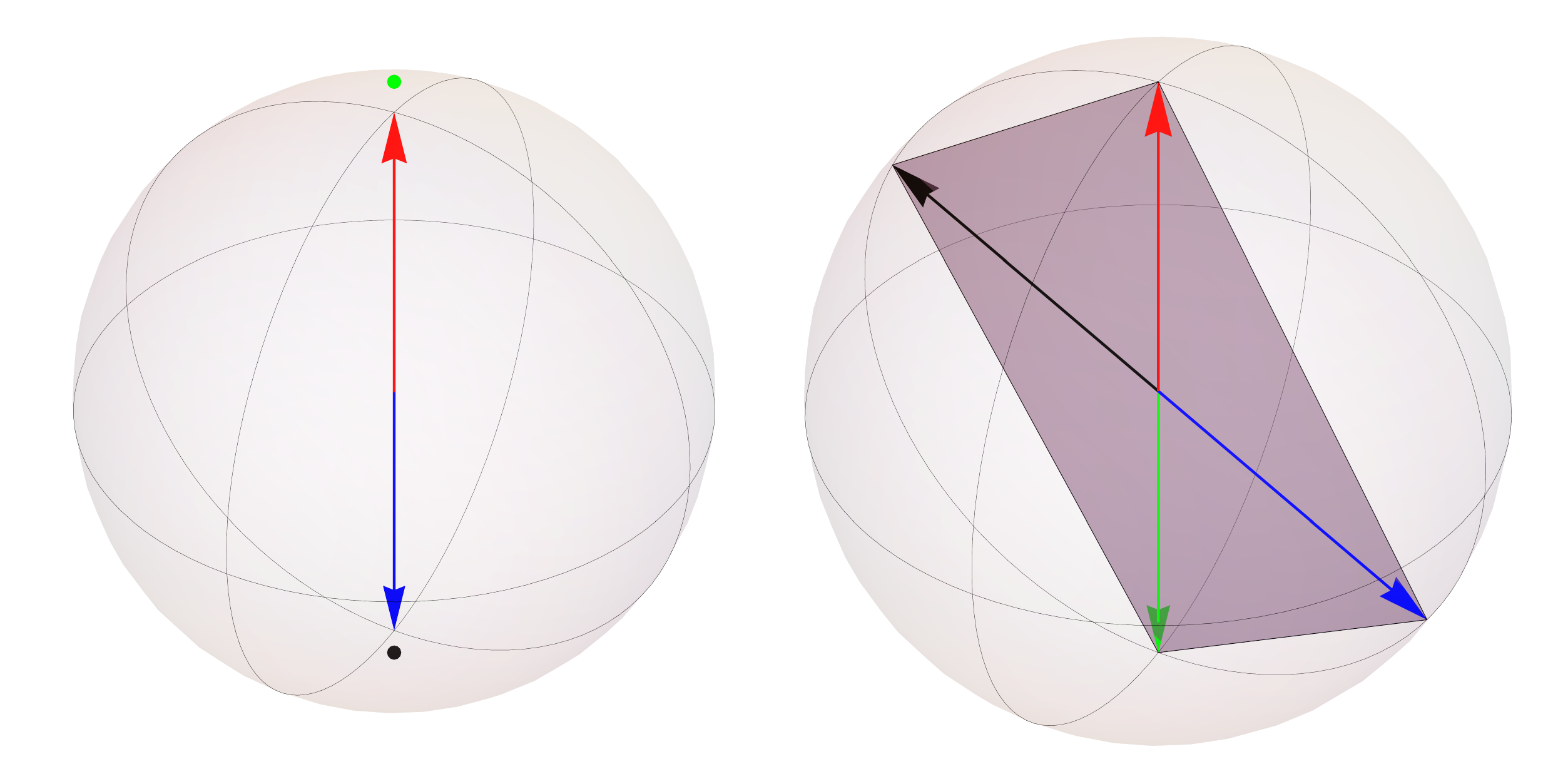}
    \caption{Both reductions density matrices for four pure states for an exemplary member of the skewed product family. Note that in the first reduction (left) all states lie on the $z$ axis, while in the second they form a rectangle in the $x-z$ plane.}
    \label{fig:Family_1}
\end{figure}

From the point of view of the Bloch ball (See Fig.~\ref{fig:Family_1}) this family is composed always from two twice degenerate points on the north and south poles in one reduction, and two pairs of opposite poles in the other.

\break


\noindent\textbf{2. \ Elegant family}
\vspace{0.2cm}


Condition (ii) yields a family which can be parametrized as

    \beqa\label{elegantfamily}
    {\scriptsize
    I^{(2)}_{\cb}=
    \frac{1}{\sqrt{2}}\left(
    \begin{array}{cccc}
    	0& 0& -e^{i\zeta}& e^{i\zeta}\\
    	c\theta&c\theta&-s\theta&-s\theta\\
    	s\theta&s\theta&c\theta&c\theta\\
      1&-1&0&0
    \end{array}
    \right),}
    \eeqa 
where we have introduced the local transformation of  the form $\exp(i \zeta\sigma_z)^{\otimes 2}$, with $\zeta = \gamma - \beta$. Hence, this family has only 2 parameters. Note that in this case, the skewed basis and the  computational one correspond, i.e., $\ket{\varphi}=\ket{0}$. 
The squared concurrence reads:
\beq
\label{xi2}
\xi^{(2)}=\frac{\sin^2(2\theta)}{4},
\eeq
which depends only on one parameter. Note that the degree of entanglement is bound, $\xi^{(2)}\in\left[0,\frac{1}{4}\right]$. Nil entanglement (i.e., $\xi^{(2)}=0$) corresponds to $\theta=0$, which leads again to the separable basis~\eqref{skewedB}. The maximal amount of entanglement, $\xi^{(2)}=1/4$, is obtained by $\theta=\pi/4$. Note that this family contains the \textit{Elegant Joint Measurement} (EJM), which play a special role in network nonlocality ~\cite{gisin2019entanglement}.
EJM has, in fact, $\xi=1/4$, and is retrieved for $\zeta = \pi/2$. 
Since EJM is the extremal case of this family, we name this ``Elegant family". Fixing the maximal amount of entanglement, however, does not single out EJM and leads to a 1-parameter subfamily.

\begin{figure}[h]
    \centering
    \includegraphics[width=.4\textwidth]{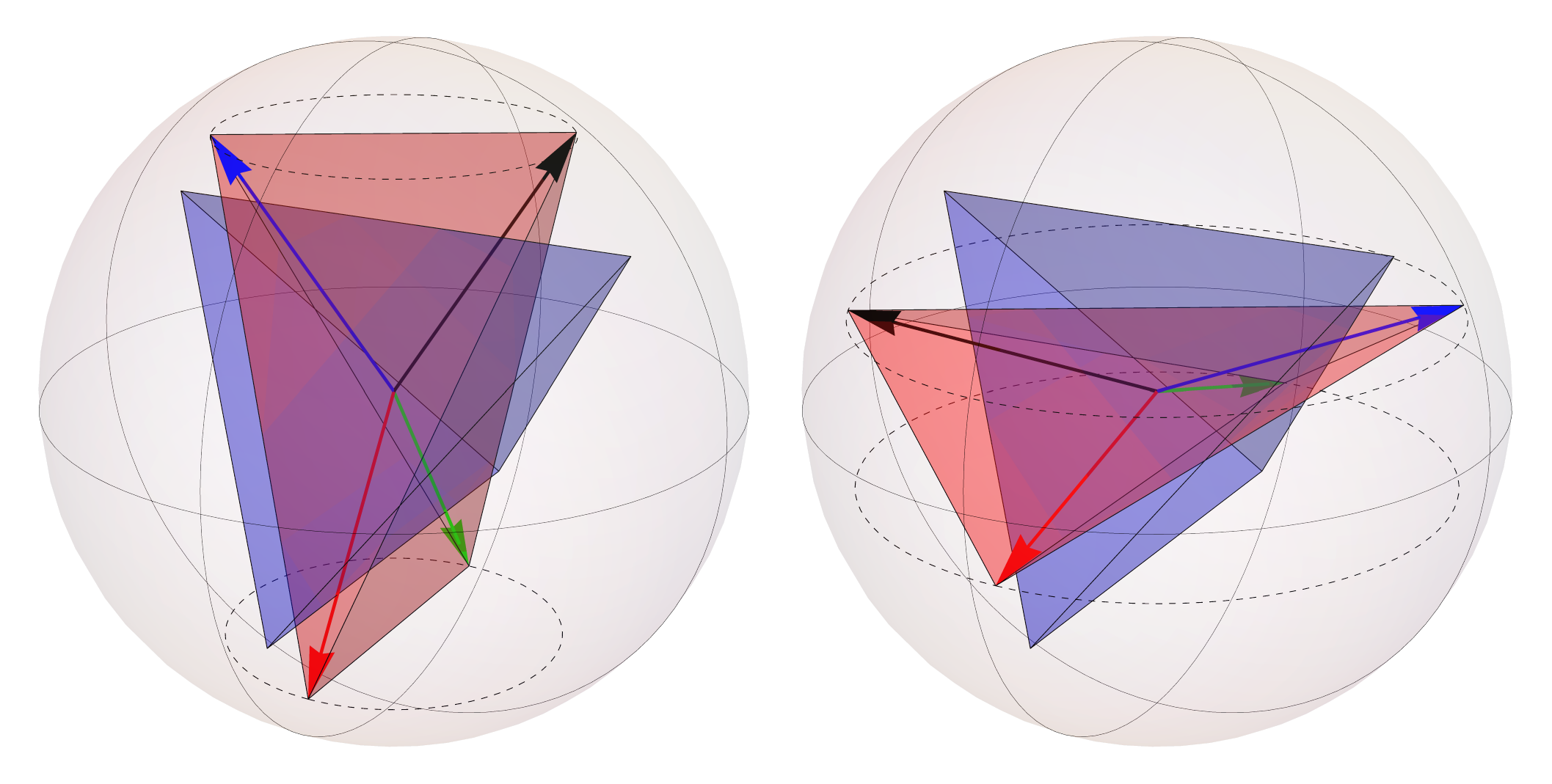}
    \caption{Partial traces for a selected member of the Elegant family (in red) together with the EJM (blue). A generic member of this family forms simplex structures with all states lying on cones with opening angles $2\theta$ and $\pi - 2\theta$, for the two reductions respectively; the second pair is rotated with respect to the first by an angle $\zeta$. In particular, EJM is found by setting $\theta = \pi/4$ and $\zeta = \pi/2$, thus forming two regular simplices.}
    \label{fig:Family_2}
\end{figure}

In Bloch ball representation (see Fig.~\ref{fig:Family_2}), the first two states lie on a hyperbole in the $x$-$z$ plane, whereas the other two lie on a full rotational hyperboloid with symmetry around $z$ axis. The opening angle of the limiting cone of these hyperboloids is $\theta$ in one, and $\pi - \theta$ in the other reduction. A generic member of this family forms a simplex with three pairs of edges of different lengths. The EJM is singled out by maximizing the volume of both reductions.


\noindent\textbf{3. \ Bell family}
\vspace{0.2cm}


This family is the conflation of conditions (iii.a) and (iii.b), 
which are equivalent up to a swap $\textrm{S}_{A\leftrightarrow B}$ and substituting $\tau$ by $\theta$, respectively. In both cases the phase $e^{i\beta}$ can be reabsorbed into the computational state $\ket{1}$ of the first qubit and defining $\zeta=\gamma+\beta$, or  into $\ket{0}$ of the second qubit and defining $\zeta'=\gamma-\beta$, respectively. Hence, this  family reads:
\scriptsize
\beqa\label{bellfamily}
I_{\cb}^{(3)}=
\left(
\begin{array}{cccc}
	0& 0& -c\delta\cdot e^{i\zeta}& s\delta\cdot e^{i\zeta}\\
	s\delta&c\delta&0&0\\
	s\tau\cdot c\delta&-s\tau\cdot s\delta&c\tau\cdot s\delta&c\tau\cdot c\delta\\
-c\tau\cdot	c\delta&c\tau\cdot s\delta&s\tau\cdot s\delta&s\tau\cdot c\delta
\end{array}
\right).
\eeqa 
\normalsize
Hence, one has the 3 expected parameters ($\delta,\zeta$ and~$\tau$). The tangle reads:
\beq\label{xi3}
\xi^{(3)}=\sin[2](2\delta)\sin[2](\tau),
\eeq
which depends on 2 parameters and varies between $0$ and $1$. For   $\delta=\pi/4$ and $\tau=\pi/2$ one achieves maximally entangled states, i.e., $\xi^{(3)}=1$. This is equivalent to the standard \textit{Bell State Measurement} (BSM),
which is the unique maximally entangled basis up to local transformations~\cite{Popescu1994causality}.  This  thus suggests the name of this family. 
For nil entanglement, i.e., $\xi^{(3)}=0$, one has either $\delta=0$, or $\cos(\tau)=\pm1$; both cases are equivalent to the already discussed separable family~\eqref{skewedB}.

\begin{figure}[h!]
    \centering
    \includegraphics[width=.4\textwidth]{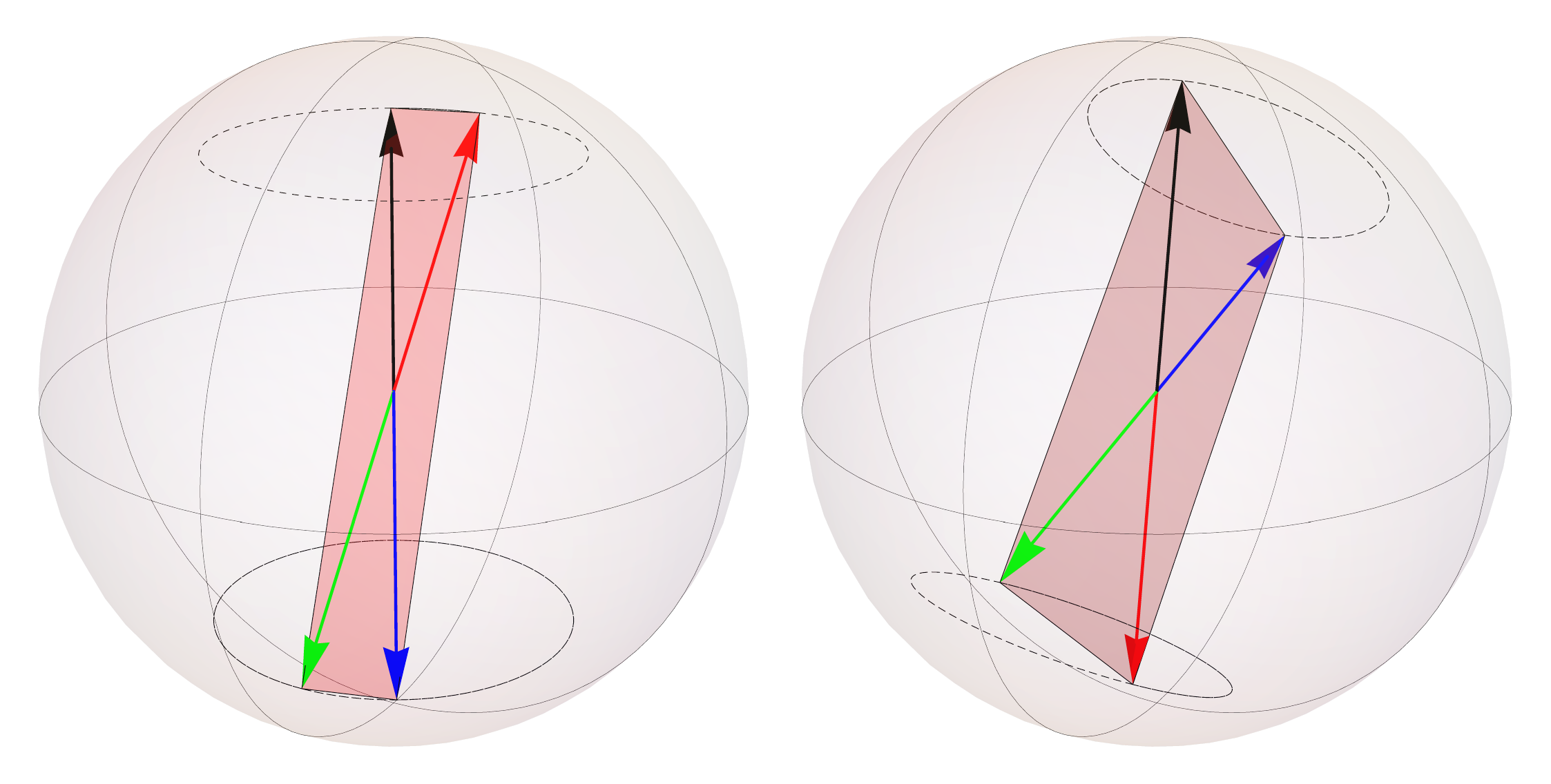}
    \caption{Generic member of the Bell family. Note that the four vectors in both reductions form two rectangles, with the first one lying on a cone with the rotation axis along $z$ axis.}
    \label{fig:Family_3}
\end{figure}

Despite full range of attainable entanglement, from the perspective of the Bloch ball, this family always produces rectangles lying in the $x$-$z$ plane in one reduction, and in a rotated plane in the other, with rotation being controlled by $\zeta$ phase.
In particular, we note that in Bloch representation (see Fig.~\ref{fig:Family_3}) a part of the Bell family will overlap with the subset of Elegant family with $\zeta = 0$. 

An alternative derivation of this family, resulting in a canonical form, is given in SM, section B.
\vspace{0.2cm}

\break
\noindent\textbf{4. \ General family}
\vspace{0.2cm}


In the case of condition (iv) we find that necessarily we have $e^{i\gamma}=-e^{\pm i\beta}$
, which yields the relation:
\beq\label{para}
\tan(\tau)=\frac{\cos(2\delta)\sin(\theta)}{\sin(2\delta)\cos(\beta)}.
\eeq
Hence, the parametrization reads:
\scriptsize
\beqa\label{monster}
I^{(4)}_{\sb}=
\left(
\begin{array}{cccc}
	0& 0& c\delta\cdot e^{\pm i\beta}& -s\delta\cdot e^{\pm i\beta}\\
	s\delta\cdot c\theta&c\delta\cdot c\theta&-s\delta\cdot s\theta&-c\delta\cdot s\theta\\
	s\delta\cdot s\theta&c\delta\cdot s\theta&s\delta\cdot c\theta&c\delta\cdot c\theta\\
	-c\delta\cdot e^{i\beta}&s\delta\cdot e^{i\beta}&0&0
\end{array}
\right).
\eeqa 
\normalsize
The expected 3 parameters are $\delta,\theta$ and $\beta$.
The tangle reads:
\beqa\label{xi4}
\xi^{(4)}&=&\frac{\sin[2](2\theta)\sin[2](2\delta)}{4}\cdot \nonumber\\
&&\cdot\frac{\sin[2](2\delta)\cos[2](\beta)+\cos[2](2\delta)}{\sin[2](2\delta)\cos[2](\beta)+\cos[2](2\delta)\sin[2](\theta)},
\eeqa
which varies between $0$ and $1$. Note that this is the most general family of iso-entangled bases, for its degree of entanglement depends on all the three parameters and it has overlaps with all the other families. Furthermore, a generic basis from this family will yield non-degenerate simplices in both reductions.

Note that Eqs.~\eqref{para} and~\eqref{xi4} have five singularity points. Studying the (directional) limits of these multivariable functions yields the following cases:
\begin{itemize}
 
  \item  $\lim \beta \rightarrow \pi/2$ reduces the General family~\eqref{monster} to a two-parameter subfamily of the Bell family~\eqref{bellfamily}. In particular, this implies that $\tau \rightarrow{} \pi /2$ and $\xi^{(4)} \rightarrow \cos[2](\theta)\sin[2](2\delta)$, which has the same form of Eq.~\eqref{xi3}. Note, that the Bell family depends on the same number of parameters as the General family, therefore it cannot be fully retrieved by any of the limits.

  \item $\lim \delta \rightarrow \pi/4$ reduces the General family to the Elegant family ~\eqref{elegantfamily}. 

  \item   $\lim \theta \rightarrow 0$ and $\delta \rightarrow 0$ reduces General family to the skewed product family~\eqref{separablefamily}, independently of the direction of approach of these limits.

  \item $\lim \beta \rightarrow \pi/2$ and $\delta \rightarrow \pi/4$ leads to an interpolationbetween a part of the Elegant family and a subfamily of the Bell family, depending on the direction of approach of the limit. In Ref.~\cite{tavakoli2021bilocal}, a one-parameter iso-entangled family was proposed that also interpolates between EJM and BSM. However, this cannot be contained within this limit case because the latter does not admit regular-simplices within the reductions, contrarily to the family in~\cite{tavakoli2021bilocal} (see SM, section C).

  \item $\lim \beta \rightarrow \pi/2$ and $\lim \theta \rightarrow 0$ leads to a subfamily of the Bell family wherein, however, the degree of entanglement is upper bounded, with the bound depending on the angle of approach $\phi\in(-\pi/2,\pi/2)$ as $(1+\tan(\left|\phi\right|))^{-2}$. 
\end{itemize}

From this, one sees that the three particular families~\eqref{separablefamily},~\eqref{elegantfamily}, and~\eqref{bellfamily} (partly) form the closure of the General family.

\section{An application to quantum networks}


Let us consider a triangular network scenario, in which Alice, Bob and Charlie share pairwise a Bell state of two qubits, e.g.,
\begin{equation} \label{eq:Network_state}
    \ket{\Psi}_{ABC} = \ket{\psi_+}_{AB}\otimes\ket{\psi_+}_{AC}\otimes\ket{\psi_+}_{BC},
\end{equation}
and each chooses a basis to perform a joint measurement on their pair of qubits (see Ref.~\cite{renou2019genuine}).
The scenario is said to be Output-Permutation Invariant (OPI) if the output probability distribution can be defined by three constants
\begin{align}
    p_1 & = p_{iii}, &
    p_2 & = p_{\sigma(iij)}, &
    p_3 & = p_{\sigma(ijk)},
\end{align}
for $i\neq j\neq k\neq i$ and any permutation $\sigma$; intuitively, it means that no node, nor output, of the network is distinguished. Similar notion can be defined for larger networks based on the network graph automorphism group. 


\begin{figure}[h]
    \centering
    \includegraphics[width=.48\textwidth]{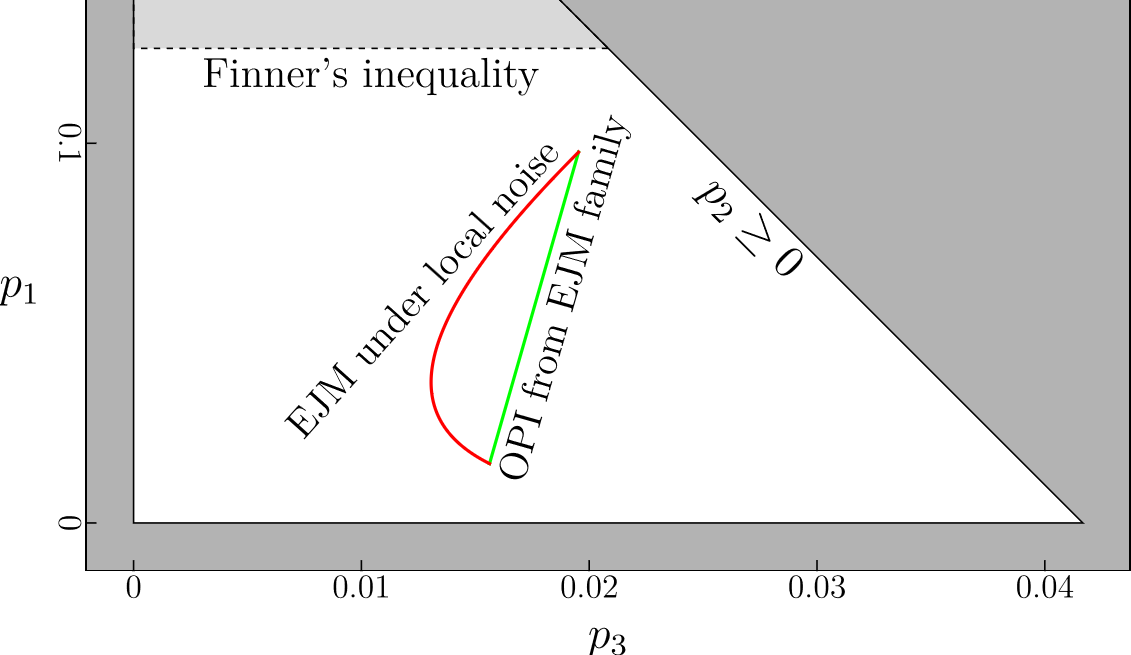}
    \caption{$p_3$-$p_1$ plane for OPI measurements, with the red line representing probabilities corresponding to EJM acting on Bell states under local noise, $\Phi_\epsilon(\op{\psi_+})^{\otimes3}$, while green line corresponds to the OPI stemming from the Elegant family acting on the network state $\op{\Psi}$ from~\eqref{eq:Network_state}. The Finner inequality is known to be a bound for local and quantum distributions~\cite{renoucorr}.}
    \label{fig:my_label}
\end{figure}

Since the iso-entangled bases set each of the measurement states on equal footing, they appear to be natural candidates for measurements realizing OPI in such networks. We find that setting $\beta = \gamma + \pi/2$ and then $\gamma = \frac{1}{2}\arccos(-\sin(2\theta))$ in the Elegant family leads to a 1-parameter subset of measurements which leads to OPI distributions. Interestingly, none of the measurements in this family, except for the extremal points, remains OPI under local noise $\Phi_\epsilon(\rho) = (1-\epsilon)\rho + \frac{\epsilon}{4}\mathbb{I}$ acting on each edge of the network (see Fig. \ref{fig:my_label}).




\section{Discussion and outlook}
In this letter, we have provided complete classification of all the equivalence classes of bases of two qubits, whose four states have all the same degree of entanglement (i.e., iso-entangled bases). In particular, we have shown that there exist four inequivalent families of equivalence classes, characterized by their numbers of parameters and geometrical constraints of their reductions in the Bloch ball representation.

This study represents a first necessary step towards a deeper  understanding of entangled measurements, a topic that has received surprisingly little attention---especially if compared to entangled states between distant parties---despite their pivotal importance in quantum computation, and other quantum tasks (such as quantum teleportation, dense coding, or the activation of nonlocality in networks). 

Although our findings provide the theoretical framework for further  studies, many questions remain open. Most of the aforementioned tasks, such as quantum teleportation or dense coding, make use of Bell State Measurements.
Our work provides the tool to start asking in systematic manner questions like: 
for which tasks partially entangled measurements provide stronger resource than the maximally entangled ones?
This can bring novel insights into nonlocality, especially in the context of quantum networks with no inputs, in which nonlocality is triggered exclusively by the selected measurements.
Moreover, further questions arise concerning implementability:
which of the entangled measurements can be experimentally realised using standard resources such as linear optical elements?

Furthermore, this preliminary study has addressed only the problem of entangled measurements in the simplest case of two qubits. The natural extension to higher dimensions turns out to be hard, with sparse known examples in literature~\cite{werner2001all, rajchel2018robust, karimipour2006equientangled, Czartowski2021bipartitequantum}. In SM, section D, we provide a short review of already known families together with a new family of partially entangled bases. This represents a first attempt towards a generalization to higher dimensions that will remain as a direction of future research.




\section*{Acknowledgements}
We thank Otfried G\"uhne for pointing out Ref.~\cite{sanpera1998local} to us. F.D.S. acknowledges support from FWF (Austrian Science Fund) through an Erwin Schr\"odinger Fellowship (Project J 4699s). J.Cz. and K.Ż. gratefully acknowledge financial support by Narodowe Centrum Nauki under the Quantera project number 2021/03/Y/ST2/00193 and the project number 2019/35/O/ST2/01049. The research has also been supported by a grant from the Priority Research Area DigiWorld under the Strategic Programme Excellence Initiative at Jagiellonian University. N.G. acknowledges support from the Swiss National Science Foundation via the NCCR-SwissMap.

\bibliography{References}

\clearpage

\renewcommand{\theequation}{SM\arabic{equation}}
\setcounter{equation}{0}



\section*{SM A -- canonical local form of the basis}
Let us denote with $B$ the matrix whose columns are the vectors $\{\ket{\psi_1},\ket{\psi_2},\ket{\psi_3},\ket{\psi_4}\}$ with coefficients expressed in the computational basis. Any 2-dimensional subspace spanned by, e.g., $\ket{\psi_3}$ and $\ket{\psi_4}$, necessarily contains (at least) one product state (Theorem 2 in Ref. \cite{sanpera1998local}). 
We may introduce local rotations in such a way that it corresponds to the state
$\ket{0,0}$. Hence, the first line of the matrix representing that basis starts with 2 zeros. Similarly, the subspace spanned by $\ket{\psi_1}$ and $\ket{\psi_2}$ contains a product state $\ket{\vartheta,\varphi}$ which is necessarily orthogonal to $\ket{0,0}$. Without loss of generality, let us assume $\ket{\vartheta}=\ket{1}$, whereas $\ket{\varphi}=\cos(\tau)\ket{0}+\sin(\tau)\ket{1}$ and $\ket{\varphi^\perp}=\cos(\tau)\ket{1} - \sin(\tau)\ket{0}$. Hence, one gets $\psi_{33}\cdot\psi_{44}=\psi_{34}\cdot\psi_{43}$, where $\psi_{jk}=\chi_{jk}e^{i\phi_{jk}}$ denotes the element $j,k$ of the matrix $B$, i.e., the $k$th component of vector $\ket{\psi_j}$ in the computational basis. Next, we choose the phases of $\psi_3$ and $\psi_4$ such that their 4th components are real, i.e., $\psi_{34}=\chi_{34}$ and $\psi_{44}=\chi_{44}$. The aforementioned relation implies that $\phi_{33}=\phi_{43}:=\phi$. Thus, it is possible to remove this phase by applying the local transformation of the form $\operatorname{exp}(i(\sigma_z\otimes\opone-\opone\otimes\sigma_z)\phi)$. Hence, the general form of our basis
written in the computational basis is given by
\beqa \label{basis1}
B_{\ket{0}}=
\left(
\begin{array}{cccc}
	0& 0& \cdot& \cdot\\
	\cdot&\cdot&\cdot&\cdot\\
	\cdot&\cdot&\chi_{33}&\chi_{43}\\
	\cdot&\cdot&\chi_{34}&\chi_{44}
\end{array}
\right),
\eeqa 
where the 4 displayed entries are real and satisfy the relation $\chi_{33}\cdot\chi_{44}=\chi_{34}\cdot\chi_{43}$.

Let us now write  B in the skewed basis provided in Eq.~(3)
. This implies that the last 2 entrees of the last line are also zeros. Starting from the form of the basis provided in Eq. (19)
, let us choose the global phases such that the third component of each vector is real, i.e., $\psi_{k3}=\chi_{k3}$. Applying the local change of basis in Eq. (3)
, leads to 
\beqa
\label{onbasis2}
B_{\ket{\varphi}}=
\left(
\begin{array}{cccc}
	0& 0& \psi_{31} & \psi_{41}\\
	\psi_{12}&\psi_{22}&\psi_{32}&\psi_{42}\\
	\chi_{13}&\chi_{23}&\chi_{33}&\chi_{43}\\
	\psi_{14}&\psi_{24}&0&0
\end{array}
\right).
\eeqa 
These coefficients depend at this point, in general, on 20 real parameters (21 counting also $\tau$ in the definition of the skewed basis). 

We now impose the orthogonality constraints between the pair of vectors for which two components of the scalar product vanish. Denoting by $\chi_{jk}$ the norm of each (in general complex) element $\psi_{jk}$, from $\braket{\psi_1}{\psi_3}$ it follows that  $\chi_{12}\cdot\chi_{32}=\pm \chi_{13}\cdot\chi_{33}$ and $\phi_{12}=\phi_{22}(+k \pi)$; from $\braket{\psi_2}{\psi_3}$, that $\chi_{22}\cdot\chi_{32}=\pm \chi_{23}\cdot\chi_{33}$ and $\phi_{22}=\phi_{32}(+l \pi)$; from $\braket{\psi_1}{\psi_4}$, that $\chi_{12}\cdot\chi_{42}=\pm \chi_{13}\cdot\chi_{43}$ and $\phi_{12}=\phi_{42}(+m \pi)$; and finally, from $\braket{\psi_2}{\psi_4}$, one finds that $\chi_{22}\cdot\chi_{42}=\pm \chi_{23}\cdot\chi_{43}$ and $\phi_{22}=\phi_{42}(+n \pi)$. 

In the next step we reintroduce the normalisation constraints.
Note that since in the considered  skewed basis, each vector has only 3 non-zero components, such that the normalization can be written as $|\psi_j|^2=a_j^2+b_j^2+c_j^2$, where each $a_j,b_j,c_j$ is one of the non-zero real component $\chi_{jk}$ of the $j-$th vector. Note that imposing the normalization conditions $|\psi_j|^2=1$ leads to the following general parametrization:
\beqa \nonumber
\begin{array}{cc}
    a_j=\sin(\alpha_j) \sin(\beta_j) \\
     b_j= \sin(\alpha_j)    \cos(\beta_j)\\
 c_j=\cos(\alpha_j).
\end{array}
\eeqa
This, together with the four orthogonality conditions above, leads to reduce the matrix of coefficients to 8 real parameters (9 counting also $\tau$ in the definition of the skewed basis). After this simplification, we impose the orthogonality contraints of the last two pair of vectors $\braket{\psi_1}{\psi_2}$, and $\braket{\psi_3}{\psi_4}$, which remove two additional parameters. Finally, one can apply a further local rotation that eliminates one last parameter, leading to the form in Eq. (4) 
which depends, as anticipated, on 5 explicit parameters (6 if one counts also the implicit $\tau$).


\section*{SM B -- canonical form of the Bell family}

Here we provide a canonical form of the Bell family by starting from its geometric structure, which will allow us to give the three free parameters explicit interpretations. We start by defining local bases
\begin{equation}
    {\small
    \begin{aligned}
        \ket{A_1} & = \cos x\ket{0}\!+\!\sin x\ket{1} &
        \ket{A_2} & = \cos x\ket{0}\!-\!\sin x\ket{1} \\
        \ket{B_1} & = \cos y\ket{0}\!+\!\sin y\ket{1} &
        \ket{B_2} & = \cos y\ket{0}\!-\!\sin y\ket{1}
    \end{aligned}}
\end{equation}
together with the orthogonal states marked by the upper index, $\ket{\cdot^{\perp}}$, defined such that the second entry is dephased. Then, we define the four bipartite states in the basis as

\begin{equation}
    \begin{aligned}
        \ket{\psi_1} & = \cos z\ket{A_1}\ket{B_1} + e^{i\phi_1} \sin z\ket{A_1^\perp}\ket{B_1^\perp} \\
        \ket{\psi_2} & = \cos z\ket{A_1^\perp}\ket{B_2} + e^{i\phi_2} \sin z\ket{A_1}\ket{B_2^\perp} \\
        \ket{\psi_3} & = \cos z\ket{A_2}\ket{B_1^\perp} + e^{i\phi_3} \sin z\ket{A_2^\perp}\ket{B_1} \\
        \ket{\psi_4} & = \cos z\ket{A_2^\perp}\ket{B_2^\perp} + e^{i\phi_4} \sin z\ket{A_2}\ket{B_2}
    \end{aligned}
\end{equation}
with $z\in[0,\pi/4]$. These four states follow the general geometric property of Bell family considered in the Bloch ball: in both reductions there are two pairs of co-linear states, and if a pair is co-linear in reduction $A$, it is not co-linear in reduction $B$.

In order to derive generic member of the Bell family we impose that $\sin x \neq 0$ and analogically for cosines and $y$ variable. After carrying out elementary inner products between the states we find that the phases are given by
\begin{equation}
    \phi_1 = -\phi_2 = -\phi_3 = \phi_4
\end{equation}
and
\begin{equation}
    \cos(\pm\phi_1)= -\frac{\tan 2x \tan 2y}{\sin 2z}.
\end{equation}

In this way we arrive at the
canonical form for the Bell family, since the three parameters are well connected to the properties of the resulting bases -- $x$ and $y$ angles are connected to the geometric arrangement in the Bloch ball, while $z$ angle corresponds to the degree of entanglement in the basis. 


\section*{SM C -- placing family from \cite{tavakoli2021bilocal} within the General family}

 A family of isoentangled bases, $I^{(5)} = \left\{\ket{\psi_i}\right\}_{i=1}^4$, has been considered in \cite{tavakoli2021bilocal} in the context of violating bilocality in linear three-partite network. It can be given explicitly in the form

\begin{equation} \label{eq:CyrilFamily}
    {\footnotesize I^{(5)}_{\ket{0}} \!=\! \frac{1}{2\sqrt{2}}\!\left(\begin{array}{cccc}
        1\!+\!i & 1\!-\!i & 1\!-\!i & 1\!+\!i \\
        \!-\!i e^{i \phi }\!-\!i & i e^{i \phi }\!-i & i\!-\!i e^{i \phi } & i e^{i \phi }\!+\!i \\
        i e^{i \phi }\!-\!i & \!-i e^{i \phi }\!-\!i & i e^{i \phi }\!+\!i & i\!-\!i e^{i \phi } \\
        1\!-\!i & 1\!+\!i & 1\!+\!i & 1\!-\!i \\
    \end{array}\right)}
\end{equation}
which, \textit{a priori}, does not correspond to any of the families introduced in this work. By considering the simple fact that it interpolates between elegant joint measurement and Bell state measurement for $\phi = 0$ and $\phi = \frac{\pi}{2}$, it cannot lie in skewed-product family, Bell family or Elegant family -- therefore it is natural to assume it to be fully embeddable within the General family.

\begin{figure}[h]
    \centering
    \includegraphics[width=.5\textwidth]{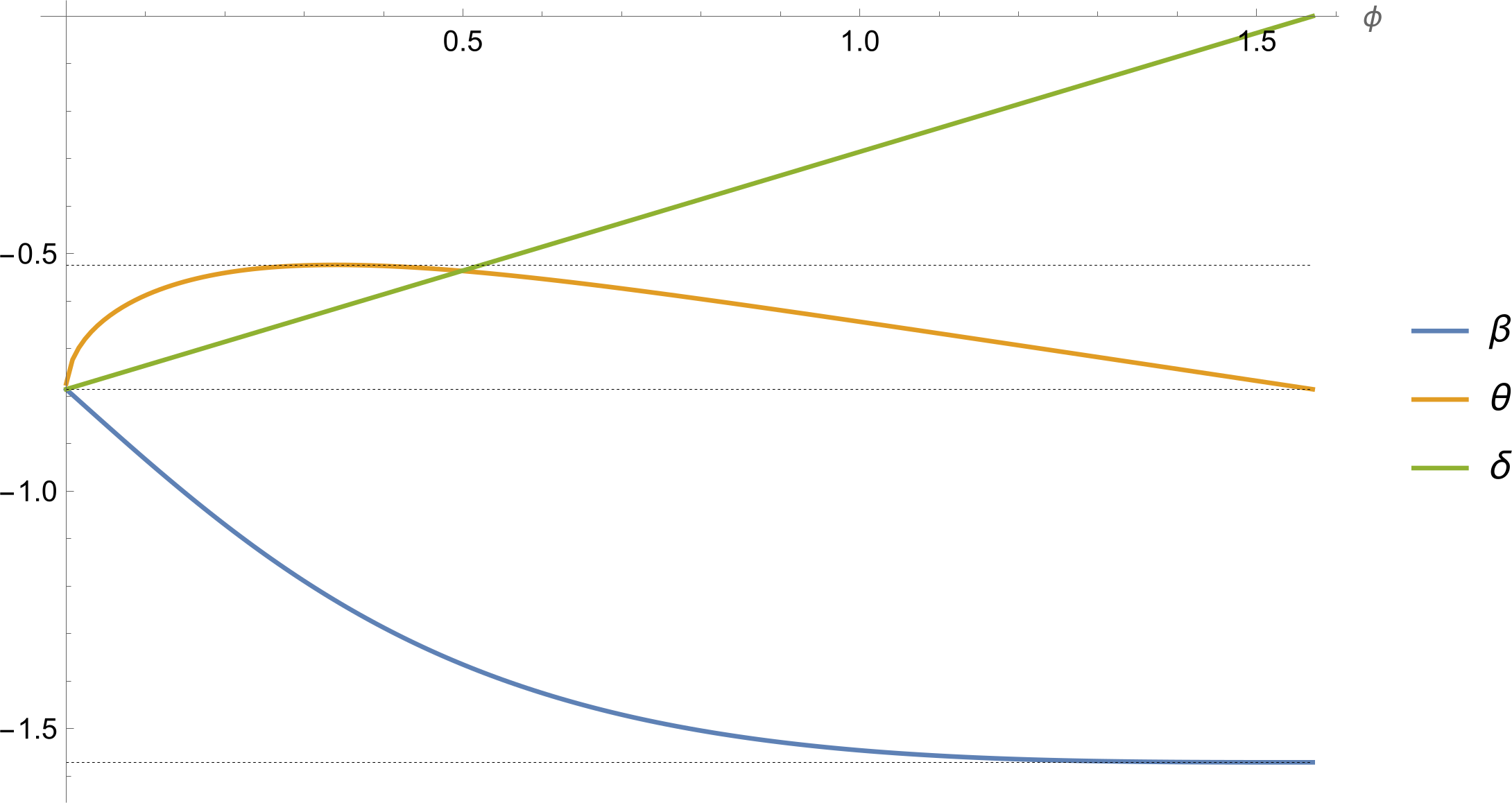}
    \caption{Family of iso-entangled bases defined in \eqref{eq:CyrilFamily} can be placed within the General family as defined in \eqref{monster} using numerically generated functions $\beta(\phi),\,\theta(\phi)$ and $\delta(\phi)$, with the last one expressible directly as $\delta = (\phi - \pi/2)/2$.}
    \label{fig:cyril}
\end{figure}

In order to find its relation to the members of the General family, let us first define $v^A_i$ as the Bloch vector corresponding to the $A$ reduction of the state $\ket{\psi_i}$, and likewise for $B$ reduction. We will consider the Gram matrices
\begin{align}
    (G_A)_{ij} = v^A_i\cdot v^A_j &&
    (G_B)_{ij} = v^B_i\cdot v^B_j
\end{align}
Similarly, we define $\tilde{G}_A$ and $\tilde{G}_B$ for members of the General family $I^{(4)} = \left\{\ket{\tilde{\psi}_i}\right\}_{i=1}^4$. Using the above, we define a cost function
\begin{equation}
     {\footnotesize F(\beta,\,\theta,\,\delta) \!=\! \sum_{i,j}\left(G^A_{ij}\!-\! \tilde{G}^A_{\sigma(i)\sigma(j)}\right)^2\!\!\!+\! \left(G^B_{ij} \!-\! \tilde{G}^B_{\sigma(i)\sigma(j)}\right)^2}
\end{equation}
and we use minimization of the above function with respect to the parameters $\beta,\,\theta,\,\delta$ with the acceptaince threshold of $F \leq \epsilon =  10^{-12}$. Using this we arrive numerically at curves $(\beta(\phi),\,\theta(\phi),\,\delta(\phi))$ for embedding the family $I^{(5)}_{\ket{0}}$ within the General family, as shown in Fig. \ref{fig:cyril}.

\section{SM D -- isoentangled families in higher dimensions}
\label{test}

The natural next step after the analysis of the simplest two-qubit bases is to shift to objects with larger local dimension, residing in spaces $\mathcal{C}^d\otimes\mathcal{C}^d$. Direct application of methods presented in this work does not seem to be realistic, thus we expect that new techniques would need to be developed to extend the current results beyond qubits. Nevertheless, there exist certain partial results and limited families in the literature.

Separable bases are the simplest case, and can be fully given in terms of \textit{conditional measurement bases} \begin{equation}
    \ket{\psi_{ij}} = \ket{i}\otimes\ket{j_i},
\end{equation} 
where one imposes the relations
\begin{equation}
    \braket{i}{i'} = \delta_{ii'},\,
    \braket{j_i}{j_{i'}} = \delta_{jj'}.
\end{equation}
and the product of the form $\braket{j_i}{j'_{i'}}$ needs not to be defined for $i\neq i'$.
Using local transformations we can always set the local bases $\ket{i}$ and $\ket{j_0}$ to the computational basis, thus simplifying this family to direct sum of unitaries, $\sum_{i=1}^d U_i$.

The other extreme---bases composed of maximally entangled states---have been considered by Werner \cite{werner2001all}, where a 1-to-1 equivalence between maximally entangled bases, unitary bases, teleportation schemes and dense coding schemes has been established. Furthermore, a family of \textit{shift-and-multiply} bases based on Hadamard matrices and Latin squares has been introduced therein. 

Limited families of isoentangled bases have been considered in \cite{rajchel2018robust, Czartowski2021bipartitequantum, karimipour2006equientangled}, without claims of being complete constructions.
Below we demonstrate a method to merge methods from \cite{werner2001all} and \cite{rajchel2018robust}, arriving at a new family of bases with intermediate entanglement degrees. 

First we focus on a construction from \cite{werner2001all}. Therein, a family of unitary bases based on Hadamard matrices and Latin squares is introduced. Given a $d\times d$ Latin square $\lambda(j,k):I_d\times I_d\mapsto I_d$ and $j$ (not necessarily distinct) Hadamard matrices, the set of unitary matrices $U^{ij}$ o
f dimension $d$ is defined by
\begin{equation}
    U^{ij}\ket{k} = (H^j)_{ik}\ket{\lambda(j,k)}
\end{equation}
and they provide an orthogonal unitary basis due to the easily verifiable property
\begin{equation}
    \text{tr}(U^{ij\dagger}U^{i'j'}) = d\delta_{ii'}\delta_{jj'}.
\end{equation}
It is also easily verified that vectorized versions $$U \mapsto \ket{U} = \frac{1}{\sqrt{d}}\sum_{nm} U_{nm}\ket{nm}$$ provide a maximally entangled bases of the $d^2$-dimensional Hilbert space. 

We may now replace Hadamard matrices with robust Hadamard matrices considered in \cite{rajchel2018robust}, such that $R^jR^{j\dagger} = \mathbb{I}$. The corresponding bistochastic matrix has the structure
\begin{equation}
    \left|R^j\right|^2 = \left(\begin{array}{cccc}
    a & b & \hdots & b \\
    b & a & \hdots & b \\
    \vdots & \vdots & \ddots & \vdots \\
    b & b & \hdots & a
    \end{array}\right)
\end{equation}
with absolute value understood entry-wise. Then, \textit{a non-unitary shift-and-multiply basis}
\begin{equation}
    B^{ij}\ket{k} = (R^j)_{ik}\ket{\lambda(j,k)}.
\end{equation}

Entire proof from \cite{werner2001all} can be retraced to demonstrate the orthogonality of the resulting states $\ket{B^{ij}}$ and the isoentangled character of the basis down to the specific Schmidt coefficients follows directly from the form of vectorization map.

Indeed, one can extend the above construction to all unitary $R^j$ such that all rows/columns contain the same amplitudes up to permutation.

It is important to note that the above construction is distinct from equientangled bases given in \cite{rajchel2018robust}, which is evident already on the level of number of robust Hadamards $R^j$ used in the construction.

\clearpage










\end{document}